\documentclass[sigconf]{acmart}
\usepackage{alltt                                    
          , multirow
          , booktabs
          , listings
          , graphicx
          ,float
          ,verbatim
         ,mathtools
	   ,url
	   ,amsmath
}
\usepackage{natbib}     
\usepackage{syntax}
\usepackage{algorithm}
\usepackage{enumitem}
\usepackage{threeparttable}
\usepackage{pbox}
\usepackage[font=small,skip=1pt]{caption}
\usepackage{algpseudocode}
\usepackage{framed}
\usepackage{hhline}
\usepackage{algpseudocode}
\usepackage{balance}
\usepackage{color, colortbl}
\usepackage{tikz}
\usepackage{verbatim}
\usetikzlibrary{arrows,shapes,backgrounds}

\tikzstyle{vertex}=[ellipse,fill=black!25,minimum size=20pt, inner sep=0pt]
\tikzstyle{edge} = [draw,thin,-]
\tikzstyle{glabel} = [text width=1cm,text centered,font=\bf]
\pgfdeclarelayer{bg}    
\pgfsetlayers{bg,main}  

\hypersetup{draft}

\setcopyright{rightsretained}

\usepackage{expl3}
\ExplSyntaxOn
\newcommand\latinabbrev[1]{
  \peek_meaning:NTF . {
    #1\@}%
  { \peek_catcode:NTF a {
      #1., \@ }%
    {#1., \@}}}
\ExplSyntaxOff

\algnewcommand\algorithmicswitch{\textbf{switch}}
\algnewcommand\algorithmiccase{\textbf{case}}
\algnewcommand\algorithmicassert{\texttt{assert}}
\algnewcommand\Assert[1]{\State \algorithmicassert(#1)}%
\algdef{SE}[SWITCH]{Switch}{EndSwitch}[1]{\algorithmicswitch\ #1\ \algorithmicdo}{\algorithmicend\ \algorithmicswitch}%
\algdef{SE}[CASE]{Case}{EndCase}[1]{\algorithmiccase\ #1}{\algorithmicend\ \algorithmiccase}%
\algtext*{EndSwitch}%
\algtext*{EndCase}%

\let\footnotesize\scriptsize

\algnewcommand{\LineComment}[1]{\State \(\triangleright\) #1}

\definecolor{LightGray}{rgb}{.9 .9 .9}

\newsavebox{\supbox}
\newcommand{\bsup}{\begin{lrbox}{\supbox}$\tt\scriptstyle}
\newcommand{\esup}{$\end{lrbox}{}^{\usebox{\supbox}}}
\def\eg{\latinabbrev{e.g}}
\def\ie{\latinabbrev{i.e}}

\definecolor{lightpurple}{rgb}{0.8,0.8,1}
\definecolor{codebg}{RGB}{255,255,255}
\definecolor{commentcolor}{RGB}{11,140,11}
\begin{document}
%

\title[Improving Bug Localization with Report Quality Dynamics and Query Reformulation]{Poster: Improving Bug Localization with Report Quality Dynamics and Query Reformulation}
%

\author{Mohammad Masudur Rahman}
\affiliation{
\institution{University of Saskatchewan, Canada}
}
\email{masud.rahman@usask.ca} 
\author{Chanchal K. Roy}
\affiliation{
\institution{University of Saskatchewan, Canada}
}
\email{chanchal.roy@usask.ca}

\begin{abstract}
Recent findings from a user study suggest that IR-based bug localization techniques do not perform well if the bug report lacks rich structured information such as relevant program entity names. On the contrary, excessive structured information such as stack traces in the bug report might always not be helpful for the automated bug localization. In this paper, we conduct a large empirical study using 5,500 bug reports from eight subject systems and replicating three existing studies from the literature. Our findings (1) empirically
demonstrate how quality dynamics of bug reports affect the performances of IR-based bug localization, and (2) suggest potential ways (\eg\ query reformulations) to overcome such limitations.  
\end{abstract}


\ccsdesc[500]{Software and its engineering~Software verification and validation}
\ccsdesc[500]{Software and its engineering~Software testing and debugging}
\ccsdesc[300]{Software and its engineering~Software defect analysis}
\ccsdesc[100]{Software and its engineering~Software maintenance tools}




\copyrightyear{2018} 
\acmYear{2018} 
\setcopyright{rightsretained} 
\acmConference[ICSE '18 Companion]{40th International Conference on Software Engineering Companion}{May 27-June 3, 2018}{Gothenburg, Sweden}
\acmDOI{10.1145/3183440.3195003}
\acmISBN{978-1-4503-5663-3/18/05}

\maketitle

\section{Introduction}
Information Retrieval (IR)-based bug localization techniques rely on shared terms between a bug report and the project source \cite{buglocator,stacktrace}.
They are cheap and their performances are reported to be as good as that of spectra-based techniques \cite{raobug}. Unfortunately, \citet{parninireval} has recently reported two major limitations based on a qualitative study where they involved human developers. First, IR-based techniques cannot perform well without the presence of rich structured information (\eg\ program entity names pointing to defects or failures) in the bug report. Second, they also might not perform well with a bug report that contains excessive structured information (\eg\ stack traces).
In this paper, we conduct a large empirical study, and re-investigate the above issues. In particular, we demonstrate how quality dynamics of bug reports (\ie\ prevalence of structured information or lack thereof) influence the performances of three existing IR-based techniques for bug localization \cite{buglocator,bluir,stacktrace}, and then also discuss potential solutions (\eg\ query reformulations).

\begin{figure*}[!t]
	\centering
	\includegraphics[width=6.6in ]{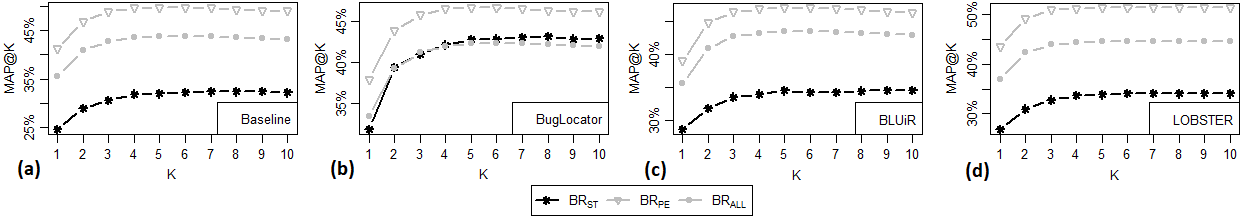}
	\caption{MAP@K of (a) Baseline, (b) BugLocator, (c) BLUiR, and (d) LOBSTER with bug reports containing stack traces (\ie\ BR$_{ST}$)}
	\label{fig:mapkrq1}
	\vspace{-.4cm}
\end{figure*}

\section{Empirical Study}\label{sec:explo}
\citet{parninireval} investigate IR-based bug localization with a user study and a limited analytical study where they replicate only one IR-based technique \cite{buglocator}.
In order to gather more strong empirical evidences, we replicate three IR-based bug localization techniques-- BugLocator \cite{buglocator}, BLUiR \cite{bluir} and LOBSTER \cite{stacktrace}, and conduct experiments using 5,500 bug reports \cite{icse2018-data} from eight open source systems. 
We answer the following research question through our study:
\FrameSep3pt
\begin{framed}
	\noindent
	\textbf{RQ:} How do existing IR-based techniques perform with the bug reports containing (a) excessive structured information (\eg\ stack traces), and (b) no structured information (\ie\ only regular texts)?
\end{framed}
\vspace{-.2cm}
\subsection{Study Design}
\label{sec:dataset}
\textbf{Dataset Collection:}
We collect RESOLVED bug reports from either BugZilla or JIRA repository of each subject system, extract \emph{changeset} (\ie\ list of changed files) from their corresponding bug-fixing commits at GitHub, and then develop bug report-solution pairs \cite{bugid}.
We also divide our report collection into three clusters--BR$_{NL}$, BR$_{ST}$ and BR$_{PE}$--based on the extent or type of structured information they have \cite{parninireval}.
Each bug report from BR$_{NL}$ contains only natural language texts whereas a report from BR$_{ST}$ contains one or more stack traces besides other structured entities. On the contrary, each bug report from BR$_{PE}$ contains regular texts, and one or more program entity names (\eg\ method names) but no stack traces.
We use a semi-automated approach in the report clustering where regular expressions and manual analysis were applied. 

\textbf{Variables of Study:}
We collect authors' implementation of BugLocator and BLUiR from their online sources, and replicate LOBSTER ourselves which is verified using
the authors' provided experimental data.
We choose three \emph{independent variables} involving bug report quality, retrieval engine and text preprocessing step and one \emph{response variable} involving bug localization performance.
We adopt a systematic approach using these variables and choose our baseline approach for bug localization.   
In short, the baseline uses \emph{whole texts} (\ie\ preprocessed without stemming) of a bug report as a \emph{query}, and \emph{Lucene} as the \emph{text retrieval engine}.


\begin{figure}[!t]
	\centering
	\includegraphics[width=2in ]{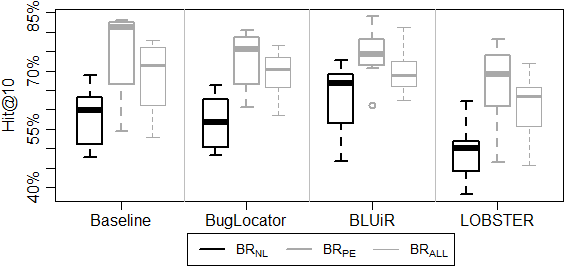}
	\caption{Hit@10 of Baseline, BugLocator, BLUiR, and LOBSTER with bug reports containing only NL texts (\ie\ BR$_{NL}$)}
	\label{fig:existingtopk}
	\vspace{-.4cm}
\end{figure}

\subsection{Results and Discussions} \label{sec:rq1}
\textbf{Impact of Excessive Structured Information:} We conduct experiments using the bug reports that contain stack traces (\ie\ BR$_{ST}$), evaluate four techniques (\ie\ three existing, one baseline), and report our findings. We found that \emph{BugLocator} performed the best among all four techniques in terms of Hit@K, Mean Reciprocal Rank@K, and Mean Average Precision@K. From Fig. \ref{fig:mapkrq1}, we see that BugLocator achieves $\approx$40\% mean average precision for various Top-K results. 
However, precision measures of each technique (especially BLUiR and LOBSTER) for BR$_{ST}$ are significantly lower (\ie\ all \emph{p-values}$<$0.05 and Cliff's 0.82$\le$$\Delta\le$1.00) than that for BR$_{PE}$ (\ie\ reports containing program entities) or BR$_{ALL}$ (\ie\ all bug reports). 
That is, the same technique performs significantly lower than the average (\ie\ for BR$_{ALL}$) when the reports contain stack traces. 
Such finding supports the conjecture about negative impacts of noisy queries on the IR-based bug localization \cite{parninireval}, and thus also warrants for query refinement prior to the bug localization.  


\textbf{Impact of the Absence of Structured Information:} We also conduct experiments with the bug reports that contain only natural language texts but no structured entities (\ie\ BR$_{NL}$), and evaluate four techniques under study. We found that \emph{BLUiR} performed the best in terms of several performance metrics. Fig. \ref{fig:existingtopk} shows the  Hit@10 of all techniques for BR$_{NL}$ dataset. It is interesting to note that the Hit@10 of BLUiR is significantly higher (\ie\ \emph{p-value}=0.02$<$0.05, $\Delta$=0.50 \emph{(large)}) than that of BugLocator although BugLocator performed the best with BR$_{ST}$ dataset. Furthermore, no existing technique except BLUiR strangely performs better than the baseline for this dataset, BR$_{NL}$, which is a bit counter-intuitive. 
Such finding might explain the role of underlying text retrieval engines given that each of the four techniques used the same query (\ie\ from BR$_{NL}$) for the bug localization. However, it also should be noted that each technique performed significantly lower (\ie\ \emph{p-values}$<$0.05 and 0.63$\le$$\Delta\le$0.84 (\emph{large})) for BR$_{NL}$ than their average performance (\ie\ for BR$_{ALL}$) irrespective of their back-end retrieval engines. That is, the queries were possibly not good enough for the localization due to their lack of relevant structured information, and thus, they need meaningful expansions.

From Figures \ref{fig:mapkrq1}, \ref{fig:existingtopk}, we also see that
no existing technique is robust enough against either collection -- BR$_{ST}$ or BR$_{NL}$. While BugLocator performs well for BR$_{ST}$, it does poorly for BR$_{NL}$. The opposite is true for BLUiR. 
On the other hand, every technique performs exceptionally well (\ie\ 70\%-80\% median Hit@10) with the high quality bug reports, \ie\ BR$_{PE}$. More interestingly, even the baseline Hit@10 is higher than that of all three competing techniques. 
All these findings above suggest that quality dynamics of a bug report is a major aspect of IR-based bug localization which was possibly overlooked by the earlier approaches. 
One could also argue that most of the existing IR-based techniques \cite{buglocator,bluir,raobug,ldabug} employ the whole report texts (\ie\ title and description) without major modifications as a \emph{query} for bug localization. Hence, their query could be either noisy due to excessive structured information (\eg\ stack traces) or insufficient due to the lack of relevant structured information (\eg\ program entity names). Thus, appropriate reformulations should be applied to the queries especially generated from BR$_{ST}$ and BR$_{NL}$ report categories before using them for bug localization with information retrieval.  
 
\section{Related Work}
Information Retrieval (IR)-based bug localization has been an active area of research for a while. Several studies employ traditional IR methods such as Latent Semantic Analysis (LSA), Latent Dirichlet allocation (LDA) \cite{ldabug} and Vector Space Model (VSM) \cite{buglocator,bluir,stacktrace} in the bug localization, and they are found to be cheap and effective \cite{raobug}. Unfortunately, recent findings \cite{parninireval} suggest two inherent limitations of the IR-based localization approaches. They are greatly affected by the (low) quality of the bug reports, \ie\ queries. Our work in this paper has empirically re-investigated such qualitative observations using a much larger dataset and has confirmed their validity.      

\section{Conclusion \& Future Work}
Existing IR-based bug localization techniques are significantly affected both by the overwhelming presence (\eg\ stack traces) and by the total absence of structured entities (\ie\ only regular texts) in the bug reports, \ie\ queries. In fact, no existing technique is robust enough against either of these two types of reports simultaneously. Future works should incorporate quality dynamics of bug reports and query reformulations into the IR-based bug localization in order to overcome the challenges outlined by this study.   

\section*{Acknowledgement}
This research is supported by the Natural Sciences and Engineering Research Council of Canada (NSERC), and by a Canada First Research Excellence Fund (CFREF) grant coordinated by the Global Institute for Food Security (GIFS) at University of Saskatchewan.

\bibliographystyle{ACM-Reference-Format}
\setlength{\bibsep}{0pt plus 0.3ex}
\bibliography{sigproc}  

\end{document}